\documentclass[structabstract]{aa} 

\usepackage{epsfig}
\usepackage{graphicx}
\usepackage{txfonts}
\usepackage{amssymb}
\usepackage{natbib}                               
\bibpunct{(}{)}{;}{a}{}{,}

\def\pl{{\sc pl}} 
\def\compps{{\sc compps}} 
\def\taut{\tau_{\rm T}}  
\def\bb{{\sc bb}} 
\def\nh{{$N_{\rm H}$}} 
\def\Integ{{\em INTEGRAL}} 
\def\rxte{{\em RXTE}} 
\def\swift{{\em Swift}} 
\def\chandra{{\em Chandra}} 
\def\xmm{{\em XMM-Newton}} 
\def\igr{IGR~J17511--3057} 
\def\igrone{IGR~J00291+5934} 
\def\be{\begin{equation}} 
\def\ee{\end{equation}} 
 
\begin{document} 
\title{Spectral and timing properties of  the accreting  X-ray millisecond pulsar IGR~J17511--3057} 
 
\author{M. Falanga\inst{1}
\and L. Kuiper\inst{2} 
\and  J. Poutanen\inst{3} 
\and D. K. Galloway\inst{4}
\and E. W. Bonning\inst{5} 
\and E. Bozzo\inst{6}
\and A. Goldwurm\inst{7,8}  
\and W. Hermsen\inst{2,9} 
\and L. Stella\inst{10}           
} 
 
\offprints{M. Falanga} 
 
\titlerunning{Accreting  X-ray millisecond pulsar IGR~J17511--3057}  
\authorrunning{Falanga et al.}  
  
\institute{International Space Science Institute (ISSI), Hallerstrasse 6, CH-3012 Bern, Switzerland 
 \email{mfalanga@issibern.ch}
\and SRON--Netherlands Institute for Space Research, Sorbonnelaan 2, 3584 CA, Utrecht, The Netherlands 
\and Astronomy Division,  Department of Physics, P.O.Box 3000, FIN-90014 University of Oulu, Finland  
\and School of Physics and School of Mathematical Sciences, Monash University, VIC 3800, Australia
\and  Department of Physics and Yale Center for Astronomy and Astrophysics, Yale University, P.O. Box 208121, New Haven, CT 06520-8121, USA 
\and ISDC, Data centre for astrophysics, University of Geneva, Chemin d'\'Ecogia 16, 1290 Versoix, Switzerland
\and Service dÕAstrophysique (SAp), IRFU/DSM/CEA-Saclay, 91191 Gif-sur-Yvette Cedex, France 
\and Unit\'e mixte de recherche Astroparticule et Cosmologie, 10 rue Alice Domon et Leonie Duquet, F-75205 Paris, France  
\and Astronomical Institute ``Anton Pannekoek'', University of Amsterdam, Science Park 904, 1098 XH, Amsterdam, The Netherlands 
\and INAF--Osservatorio Astronomico di Roma, via Frascati 33, 00040 Monteporzio Catone (Roma), Italy  
             } 
 
   \date{ } 
 
% \abstract{}{}{}{}{}  
% 5 {} token are mandatory 
  
  \abstract 
{\igr\ is the second  X-ray transient accreting millisecond pulsar discovered by \Integ. It was in outburst for about a month from September 13, 2009.} 
{We analyze the spectral and timing properties of the object as well as the characteristics of 
X-ray bursts with the aim to constrain the physical processes responsible for the X-ray production in 
this class of sources. 
}  
  {The broad-band spectrum of the persistent emission in the 0.8--300 keV energy band
  was studied using simultaneous \Integ, \rxte\ and \swift\ data obtained in September 2009.
    We also describe in the 2--120 keV energy range  the timing properties 
    such as the outburst light curve, pulse profile, pulsed fraction, pulsed emission, and time lags, 
    as well as study the properties of X-ray bursts discovered by \rxte\ and \Integ\ and the recurrence time.  }
  {The  broad-band average spectrum is well
  described by thermal Comptonization with an electron temperature of
  $kT_{\rm e}\sim 25$ keV, soft seed photons of  $kT_{\rm bb}\sim 0.6$ keV, and Thomson optical 
  depth $\taut\sim2$ in a slab geometry. During the outburst the
spectrum stays remarkably stable with plasma and soft seed photon
temperatures and scattering optical depth being constant within errors. We fitted the outburst profile with the exponential model, and 
using the disk instability model  we inferred the outer disk radius to be $(4.8-5.4) \times 10^{10}$ cm. 
The \Integ\ and  \rxte\ data reveal the X-ray  pulsation at a period of 4.08 milliseconds up to $\sim$120 keV. 
The pulsed fraction is shown to decrease from $\sim$22\% at 3
  keV to a constant pulsed fraction of $\sim$17--18\% between 7--30
  keV, and then to decrease again down to $\sim$13\% at 60 keV. The
  nearly sinusoidal pulses show soft lags 
  monotonically increasing with energy to about 0.2 ms at 10--20 keV similar 
  to those observed in other accreting pulsars. 
The short burst profiles indicate hydrogen-poor material at ignition, which suggests either that the accreted material is hydrogen-deficient, or that the CNO metallicity is up to a factor of 2 times solar. However, the variation of burst recurrence time as a function of $\dot{m}$ (inferred from the X-ray flux) is much smaller than predicted by helium-ignition models.} 
 {}
 
\maketitle

\section{Introduction} 
\label{sec:intro} 

The discovery of radio millisecond pulsars in binary systems in the
1970s lead to the prediction that neutron stars hosted in low %slower spinning 
mass X-ray binary systems (LMXB) were their progenitors
\citep{alpar82}. As a binary system evolves through phases of
accretion onto the neutron star (NS), it gains angular momentum from
the accreted material, which is sufficient to spin-up the NS to
a rotation period equilibrium in the millisecond range.  The first
confirmation that LMXBs can host rapidly rotating NSs was the discovery of 
coherent oscillations from type-I X-ray bursts \citep[see][ for a review]{sb06}. 
Later identification of SAX J1808.4--3658 as a 401 Hz pulsar 
\citep{wvdk98} led to the discovery of a new class of accreting NSs, 
accreting millisecond X-ray pulsars (AMXPs).  The first direct
measurement of spin-up of the NS during an accretion phase was
published by \citet{mfa05} in the source IGR J00291+5934,
strengthening the hypothesis of the recycling of old radio pulsars to
millisecond periods \citep[see also the reanalysis by][]{patruno10}. On the other hand, between the outbursts, long-term monitoring shows some AMXPs to
exhibit spin-down in quiescence \citep{HPC09,patruno10,papitto10b}. The spin frequencies of AMXPs lie
in the range of 180--600 Hz and orbital periods are between 40 min and
5 hr \citep[see reviews by][]{w06,p06}.

AMXPs exhibit many characteristics similar to other LMXBs. Their 
broadband spectra show soft thermal and hard Comptonized components \citep[see
reviews by][]{p06,mf08} similar to atoll sources in the low/hard state \citep{barret00}.
AMXPs also show quasi periodic oscillations and X-ray bursts \citep[see e.g.,][]{cmm03,watts05} . 
The pulse profiles are typically close to sinusoidal with modulations of 3--15\%. 
The pulse profiles are energy dependent and demonstrate 
soft time lag in the range up to 100--250 $\mu$s \citep[e.g.,][]{CMT98,Ford00,pg03,gp05,ft07}. 

The magnetic field of the NS as inferred 
from accretion models is relatively weak, in the range of $\sim
10^{8}-10^9$ G \citep{psaltis99,disalvo03}. Measurements of the 
inner disk radius from evolution of the pulse profiles and 
the spin-down of the pulsars between the outbursts allow a more accurate 
determination of the magnetic field of $1.5\times 10^8$ G in SAX J1808.4--3658  
\citep{ip09,HPC09} and of $2\times 10^8$ G in \igrone\ \citep{patruno10}. 

Pulsations are not detected during quiescence in either X-ray or
optical bands, nor has millisecond radio pulsed emission been found
\citep[see e.g.,][and references therein]{iacolina10}.  Following
outbursts, the optical counterpart has been observed to diminish in
intensity consistent with the X-ray flux decay rate implying the origin of the emission to be 
in the accretion disk \citep[see e.g.,][]{mfa05}.  Companion stars of AMXPs are highly evolved white or
brown dwarfs \citep{deloye03}. In the AMXPs with an orbital period in
the hour range and a hydrogen-rich donor brown dwarf, X-ray bursts
have been detected \citep[see e.g.,][]{galloway06}, during which burst
oscillations are nearly in phase with the coherent oscillation at the
spin period \citep{cmm03,SMS03}.
 
\subsection{The source \igr}

Among roughly one hundred LMXBs hosting a NS, \igr\ is the twelfth known to host an AMXP.
It is the second source discovered by \Integ\ during the Galactic
Bulge monitoring program \citep{baldovin09} that was found to be an
AMXP with \rxte\ \citep{markwardt09}.  The {\it INTEGRAL}-derived
source position for the new transient was $\sim20\arcmin$ away from
the known 435~Hz millisecond pulsar XTE~J1751--305, the source
position uncertainty of $2\arcmin$, as well as the subsequent
detection of a distinctly different pulse frequency of 244.8 Hz confirmed
the transient to be a new AMXP \citep{markwardt09}.  The orbital period of \igr\ is $\sim3.47$ hr, with an $a\sin(i)/c$ value of $\sim275.2$ lt-ms \citep[][and references therein]{riggio10}.
A candidate
near-infrared counterpart was found by \citet{torres09a} on September
22, 2009 with a $K_s$-band magnitude of 18.0$\pm$0.1, which had faded to
$K_s > 18.8$ (3$\sigma$ upper limit) by October 7, 2009
\citep{torres09b}. Radio upper limits of 0.16-0.18 mJy between September 16 and 25, 2009 
were set with the VLA by \citet{millerjones09}. The most accurate
position of the source was provided by the near-infrared observations
at $\alpha_{\rm J2000} = 17^{\rm h}51^{\rm m}08\fs64$ and $\delta_{\rm
J2000} = -30{\degr}57\arcmin40\farcs70$ \citep{torres09a}, consistent
with the 1$\sigma$ error of $0\farcs6$ of the \chandra/HETG position
\citep{nowak09}.

The first type-I X-ray burst from the source was detected by \swift\
\citep{bozzo09}, and the \swift\ monitoring of this source in outburst
was reported by \citet{bozzo10}, who found no evidence for
photospheric radius expansion during the three studied bursts. All
follow-up observations with different spacecrafts detected X-ray
bursts from the source. Burst oscillations at $\sim 245$ Hz were first
detected in an X-ray burst on September 14, 2009 by \rxte/PCA
\citep{watts09}; \xmm\ also detected X-ray bursts with burst oscillations
phase locked to the persistent pulsations \citep{papitto09}. A
\chandra\ observation on September 22, 2009 also detected an X-ray
burst \citep{nowak09}. The net persistent spectrum, 0.5--10 keV band, of \igr\ was found
for all the observations to be well-fit by an absorbed power-law model
with photon index $\sim 2$ and column density $N_{\rm
H}=(0.6-1.3)\times10^{22}$cm$^{-2}$, depending on the model used
\citep{bozzo09, nowak09,papitto10a}.

A lower limit on the companion mass of 0.13M$_{\odot}$ was determined
by \citet{markwardt09} using the mass function of the
system and assuming a NS mass of 1.4 M$_{\odot}$ and an inclination
angle of $90{\degr}$. This lower limit was later improved by
\citet{papitto10a}, who considered that no occultation or dips were
observed during the outburst of \igr. These authors also  performed  a
fit to the source spectrum with a reflection continuum model and
determined an inclination angle for the system in the range
$38-68{\degr}$. The corresponding limits on the companion mass in this
case are 0.15--0.23 M$_{\odot}$ (a NS of 1.4 M$_{\odot}$ was
assumed). We refer the reader to \citet{papitto10a} for further detail
regarding the nature of the companion star. An upper limit on the
source distance of $\sim10$ kpc was estimated first by
\citet{papitto10a} and $\sim7$ kpc  by \citet{altamirano10} assuming that the most
energetic type-I X-ray burst observed from \igr\ reached the Eddington
luminosity.
 
In this paper we report on \Integ\ and \rxte\ observations
of \igr, as well as a simultaneous \swift, \Integ, and
\rxte\ observation during the period MJD 55087--55117 (September 13 -- October 13, 2009). We study
the light curves, broad-band spectra, outburst spectral evolution, and timing properties of the source. The
properties of the largest set of X-ray bursts from this source are also investigated.

\section{Observations and data} 
 
\subsection{INTEGRAL} 
\label{sec:integral}  
 
The present data were obtained during the \Integ\ \citep{w03}
Target of Opportunity (ToO) observation during satellite revolution
846, starting on September 16 (MJD 55090.92557) and ending on September 19, 2009 (MJD 55093.58993), with a
total net exposure time of 206 ks.  The data reduction was performed using the standard Offline Science
Analysis (OSA\footnote{http://www.isdc.unige.ch/integral/analysis}) version 8.0 distributed by the Integral Science
Data Center \citep{c03}. The algorithms used for the spatial and
spectral analysis are described in \citet{gold03}. The observation, aimed at \igr,
consists of 70 stable pointings with a source position offset
$\lesssim 7\fdg0$ from the center of the field of view.  We analyzed data from the IBIS/ISGRI coded mask telescope
\citep{u03,lebr03} at energies between 18 and 300 keV and from the
JEM-X monitor, module 1 \citep{lund03} between 3 and 20 keV.  Note, since this observation was
in a rectangular pattern mode which consists of a square $5\times5$
pattern around the nominal target location, \igr\ was only  within
the source position offset of $<3\fdg5$ from the center of the field of view for 28 stable pointings. 
Therefore, for JEM-X with a field of view of $7\fdg5$ (diameter) at half response the effective exposure time was only 87.4 ks. 
 
We first deconvolved and analyzed separately the 70 single pointings
and then combined them into a total mosaic image in the 20--40 keV and
40--100 keV energy bands, respectively. In the mosaic, \igr\ is clearly
detected at a significance level of $\sim65\sigma$ for energies between 20--40 keV and
$\sim29\sigma$ at higher energies (40--100 keV).  The source position in
the 20--40 keV band is $\alpha_{\rm J2000} = 17^{\rm h}51^{\rm
m}08\fs74$ and $\delta_{\rm J2000} = -30{\degr}57\arcmin36\farcs9$
(error of $0\farcm55$ at the 90 per cent confidence level,
\citealt{gros03}), which is offset with respect to the near-infrared
position by $0\farcm07$ \citep{torres09a}. 

Figure \ref{fig:figima} shows part of the ISGRI field of view (significance map) around the
position of \igr\ in the 20--70 keV energy range. \igr\ is clearly detected,
together with two other nearby sources, at a high significance level
(white circles). In the same figure we also indicated with a 
yellow circle the position of the millisecond pulsar XTE~J1751-305 that
underwent a short period of enhanced X-ray activity during the last
part of the outburst decay of \igr\ (see Sect. \ref{sec:rxte} and Fig.
\ref{fig:fig1}).

To search for X-ray bursts, the ISGRI light curves are calculated from events selected according
to the detector illumination pattern for \igr. For ISGRI we used an
illumination factor threshold of 0.6 for the energy range 18--40 keV;
for JEM-X we used the event list of the whole detector in the 3--20
keV energy band.
 
\subsection{RXTE} 
\label{sec:rxte}  
  
\begin{figure}[t] 
{\hspace{-0.25cm}\psfig{figure=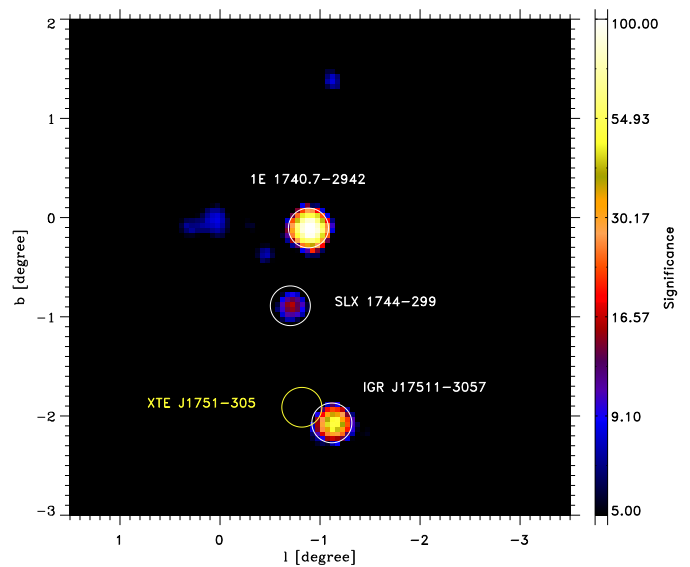,width=8.7cm}}
{\caption[]{{\it INTEGRAL}/ISGRI sky image of the field of view around \igr\
(20--70 keV). The size of each pixel in the image corresponds to 3$\arcmin$. 
It is evident from the image that the non-imaging instruments on-board \rxte\ with 
a field of view of $\sim1\degr$ (radius) would not be able to separate the
emission from the two AMXPs XTE J1751-305 and \igr\ when they are both
active in X-rays (see Fig. \ref{fig:fig1}).
\label{fig:figima}} }
\end{figure}

We used publicly available data from the proportional counter array
PCA (2--60 keV; \citet{jahoda96}) and the High Energy X-ray Timing
Experiment HEXTE (15--250 keV; \citet{rothschild98}) on-board the
\rxte\ satellite.  The FWHM of these collimator instruments is
$\sim1\degr$, and no spatial information of the photons exists.
\igr\ was monitored from September 13 to October 8, 2009 (MJD
55087.85619--55112.33693) for a total net exposure time of $\sim 455$
ks (observation ID 94041). For the light curve analysis, we also added the twelve 
pointings from observation ID 94042 ($\sim 45.2$ ks net exposure time; target XTE~J1751-305).  
In this dataset the nearby X-ray millisecond pulsar XTE~J1751-305 underwent a short outburst contemporaneous with  the latest outburst phases of \igr\ 
\citep{chenevez09,markwardt09b}.  Figure \ref{fig:figima} shows that
the angular separation between the two sources is smaller than the
field of view of the non-imaging instruments on-board \rxte\
($\sim1\degr$), and thus we were unable to disentangle the contributions
of the two objects to the total X-ray flux.

We carried out a spectral analysis, using Standard-2 data
(with 16 s time resolution) for the PCA and standard Cluster-0 data
for HEXTE.  For HEXTE, we used the on-source data, using default
screening criteria for Cluster 0. The data were extracted for the Good
Time Intervals defined by standard criteria. The PCA response matrix
was created by FTOOLS version 6.0.2 for 129 energy channels to cover
the energy range from 2--60 keV. For HEXTE we used the standard 64
energy channel response matrix for the 15--250 keV energy range. For
the timing analysis we used PCU data, collected in the {\tt
E\_125us\_64M\_0\_1s} event mode, recording event arrival times with
122.07 $\mu$s time resolution, and sorting events in 64 PHA
channels. Default selection criteria were applied.

\subsection{Swift} 
\label{sec:swift}  
 
\igr\ was also observed with \swift\ \citep{burrows05} during its outburst
in 2009. The results were first published in \citet{bozzo10}. We
used only  \swift/XRT data in window-timing mode (WT) that
were simultaneously collected with the  \Integ\ observations \citep[i.e. \swift\
observation ID 00031492005; see Table 1 in][for details]{bozzo10}. 
We refer the reader to  \citet{bozzo10} for the \swift\
XRT data reduction procedure. 

\section{Outburst light curve}
\label{sec:lc} 

Most AMXPs that underwent an outburst for a few weeks to months showed a
common outburst profile, i.e. the light curve decays exponentially
until it reaches a break, after which the flux drops linearly to the
quiescence level \citep[see e.g.,][]{gilfanov98,gp05,mfb05}. These
outburst profiles have been modeled for soft X-ray transients based on
the disk instability picture of \citet{king98}, i.e. taking into
account the disk irradiation by the central X-ray source during the
outburst. \citet{king98} showed that X-ray heating during the decay
from outburst causes the light curves of transient LMXBs to exhibit
either exponential or linear declines depending on whether or not the
luminosity is sufficient to keep the outer disk edge hot. This model
has been applied to a sample of different LMXBs, including three AMXPs
\citep{powell07}. \citet{hartman10} investigated the outburst light curve for one
of the AMXPs, IGR J00291+5934, using different models in which the decay
tail is not necessarily linear.  However, \citet{powell07} showed that
the timescale of the decay light curve and its luminosity at a
characteristic time are linked to the outer radius of the accretion
disk. A knee in the light curve of the decay from outburst of a
transient LMXB is believed to be a consequence of mass transfer onto
the outer edge of the disk, since this supply is effectively cut off
from the compact object when the outer disk enters the cool
low-viscosity state. The X-ray luminosity at which the knee occurs is
that at which the outer disk edge is just kept hot by central
illumination, allowing this radius to be calculated. In addition, the
exponential time-scale of the decay gives a second measure of the disk
radius.

\begin{figure} 
\centering 
\centerline{\epsfig{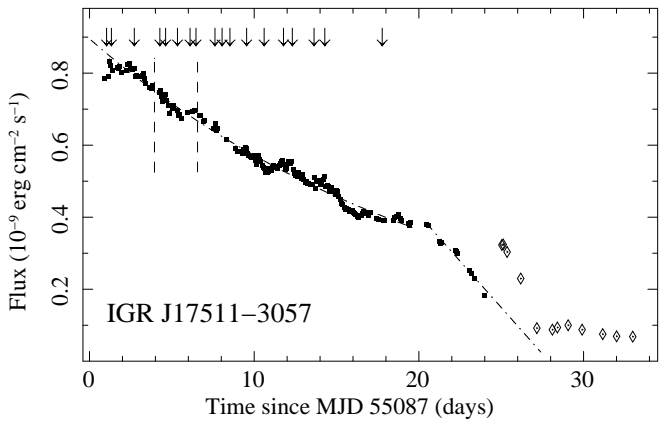} } 
\caption{\rxte/PCA (2--20 keV) outburst light curve of \igr. 
For plotting purpose we chose a bin time of 6840 s.
The count rate has been converted into flux using the
spectral results reported in Sect. \ref{sec:evolution}. The diamonds correspond to observations 
in which both XTE~J1751--305 and \igr\ were active and the instruments on-board \rxte\ were unable to disentangle the
contribution of the two sources. The arrows indicate the times of the detected
X-ray bursts (see Sect. \ref{sec:burst} and Table
\ref{table:rec}). The vertical dashed lines indicate the interval of the \Integ\ observations.  
The dot-dashed lines correspond to the best-fit exponential profile, 
$F \propto e^{-t/22.4^{\rm d}}$, and a linear decay.
%$F \propto e^{-t/5.6^{\rm d}}$. 
}
\label{fig:fig1} 
\end{figure}

The 2--20 keV band \rxte\ light curve of \igr\ has been extracted for
all the pointings and is shown in Fig. \ref{fig:fig1}, averaged over 6840 s. 
The diamonds correspond to the sum of the
emission of the X-ray millisecond pulsar XTE~J1751--305 in a weak
outburst and of \igr\ returning back to quiescence.
\rxte\ and \Integ\ detected ten and three type-I X-ray
bursts, respectively. These are indicated with arrows and are
subtracted from the light curve and discussed in Section \ref{sec:burst}. 
We converted the count rates to flux using the spectral
results from Section \ref{sec:evolution}.
The dot-dashed lines correspond to the best-fit exponential profile 
$F \propto e^{-t/22.4^{\rm d}}$ and a linear decay. 
In order to fit the light curve of the outburst decay of \igr, we excluded the \rxte\ data in which the other AMXP XTE~J1751--305 was also active. We have verified that a linear fit would be probably favoured by the data (the reduced $\chi^2$ is about half of the corresponding value for the exponential fit). However, given the relatively poor observational coverage over the last part of the outburst decay, a firm distinction between a linear and an exponential decay cannot be made.
 
From the equations (9), (23), and (21) in \citet{powell07} it is
possible to estimate the outer disk radius in two different and
independent ways. Equation (9) gives $L_X = (L_t-L_e)\,{\rm
exp}(-(t-t_t)/\tau_e) + L_e$, where $L_e$, $t_t$ (the break time),
$L_t$, and $\tau_e$ (exponential decay time) are all free
parameters. By using for these parameters the values determined from
the fit to the light curve of the outburst decay of \igr\ (see
Fig. \ref{fig:fig1}), the outer disk radius can be estimated from
$R_{disk}(\tau_e)= (\tau_e\,3\nu_{\rm KR})^{1/2} \approx
4.8\times10^{10}$ cm. Here we adopted for the viscosity near the outer disk edge, $\nu_{\rm KR}= 4\times10^{14}$
cm$^2$ s$^{-1}$, in agreement with \citet{king98} and
\citet{powell07}. An independent estimate of $R_{disk}$ can also be
obtained by using the equation (21) in \citet{powell07},
i.e. $R_{disk}(L_t)= (\Phi\,L_t)^{1/2}$. By assuming $\Phi\approx 1.3$
cm$^2$ s erg$^{-1}$ \citep[see][]{powell07}, we obtained
$R_{disk}(L_t)\approx (4.8-5.4)\times10^{10}$ cm, for a distance to
the source of 6.3 and 7 kpc, respectively (see also
Sect. \ref{sec:burst}). Note that the two independent estimates of the
inner disk radius agree remarkably well in case a distance to
the source of 6.3 kpc is considered. A similar agreement was also
found for a number of different AMXPs, see \citet{powell07}. This
radius also fulfills the condition $R_{\rm circ} < R_{\rm disk} <
b_{1}$, where $R_{\rm circ} \approx (2.8-2.3) \times 10^{10}$ cm is
the circularization radius and $b_{1} \approx (6.7-6.4)\times10^{10}$
cm is the distance of the Lagrange point $L_{1}$ from the center of
the neutron star \citep[see e.g.,][]{fr02}. To estimate these values
we used a companion star mass between 0.15 -- 0.23 M$_{\odot}$
\citep{papitto10a}, for a NS mass of 1.4 M$_{\odot}$.

\section{Spectral analysis} 
\label{sec:spectra}
  
The spectral analysis was carried out using XSPEC version 12.6
\citep{arnaud96}. For the contemporaneous data we 
 combined the 3--22 keV \rxte/PCA data, 
 and the 5--300 keV \Integ/JEM-X/ISGRI data taken on September 16--19, 2009.  
 In addition, for the low energy range 0.5--10 keV we also used the
 contemporaneous \swift\ data (ID 00031492005 WT) from \citet{bozzo10}.
For each instrument, a multiplication factor was included in the fit to take into account the
 uncertainty in the cross-calibration of the instruments. For all the
 fits the factor was fixed at 1 for the ISGRI data.  
 To follow the outburst spectra outside the \Integ\ observation interval
 we used the \rxte/PCA/HEXTE data.  To take into account contamination by the Galactic ridge emission, we used as background for the PCA spectral fitting the {\em RXTE}/PCA data collected when both pulsars were at the lowest X-ray emission level (6.3 cts/s) \citep[see also][]{papitto10a,altamirano10}. All uncertainties in the spectral parameters are given at a 90\% confidence level for
 single parameter.
   
\begin{figure} 
\centering 
\centerline{\epsfig{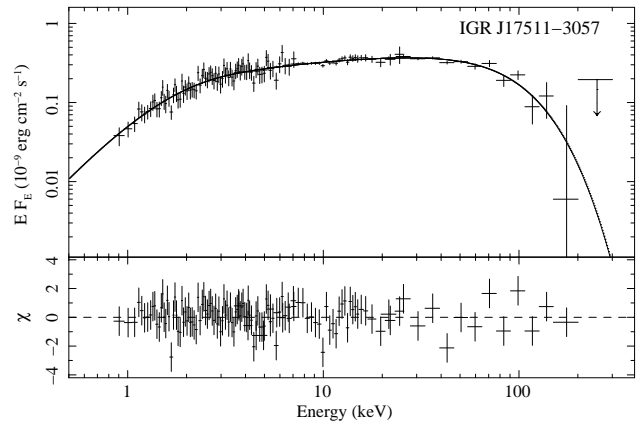} } 
\caption{Unfolded broad-band spectrum of IGR J17511-3057 fit with the thermal 
Comptonization model \compps. The data points are from XRT (0.8--7
keV), PCA (3--22 keV), and ISGRI (20-300 keV) spectra,
respectively. The total model spectrum is shown by a solid line. The lower
panel shows the residual between the data and the model.}
\label{fig:fig2} 
\end{figure} 
 
\subsection{Broad-band spectrum} 
\label{sec:spec} 
 
 We studied in detail the broad-band X-ray spectrum of \igr\ in the energy range 0.8--300 keV,
 using the joint \Integ/JEM-X/ISGRI, \rxte/PCA, and \swift/XRT data.
 For the \rxte\ data we removed the time intervals corresponding to
 the bursts.  We first fit the joint XRT/JEM-X/PCA/ISGRI (0.8--300 keV)
 spectrum using a simple photoelectrically-absorbed power-law, \pl,
 model which was found to be inadequate with a $\chi^{2}{\rm /d.o.f.} =
 417/141$.  A better fit was obtained by replacing the \pl\ with the exponentially cutoff
power-law model, {\sc cutoffpl}, ($\chi^{2}{\rm /d.o.f.} = 115/140$). We estimated a power-law photon index of $\Gamma=1.61\pm0.04$ and
 a cutoff energy of $E_{\rm c}=58\pm7$ keV.  The absorption was found to be $N_{\rm H}=0.96_{-0.09}^{+0.10} \times 10^{22} {\rm cm}^{-2}$, close to the Galactic value reported in the radio maps of \citet{dickey90}.
Adding a multi-temperature disk blackbody model, {\sc diskbb}, Êto the fit in order to include 
the contribution of the soft thermal disk emission to the total X-ray flux \citep{mitsuda84} gave only a marginal improvement to the fit ($\chi^{2}{\rm /d.o.f.} = 105/146$). For this spectral component 
we obtained an inner disk temperature of $kT_{\rm in}=0.13\pm0.02$ and an inner radius of $R_{\rm in}\sim 601 \sqrt{{\rm cos}\; i}$ km, in agreement with the results reported by \citet{bozzo10}. 
The best fit value of $R_{\rm in}$ suggests a relatively large value of the inner disk radius that Êis hardly compatible with the idea that in AMXP sources the accretion disk extends down to a region very close to the NS surface (i.e., 10 km). A better description of the emission from \igr\ at the lower energies ($<$2 keV) was obtained by using the higher spectral resolution of the instruments on-board  {\it XMM-Newton}, as discussed in \citet{papitto10a}.
     
In order to compare the \igr\ spectrum with previously observed
broad-band spectra of the same source class
\citep[e.g.,][]{gdb02,gp05,mfa05,mfb05,mfc07,ip09}, we replaced the 
cutoff power-law model with the thermal Comptonization model,
\compps, in the slab geometry \citep{ps96}. The main model parameters are the Thomson optical depth 
$\tau_{\rm T}$ across the slab, the electron temperature $kT_{\rm e}$,
the temperature $kT_{\rm bb}$ of the soft seed blackbody photons assumed to be
injected from the bottom of the slab, the emission area $A_{\rm bb}$, 
and the inclination angle $\theta$ between the slab normal and the line of sight. The best fit
parameters of all the models used to fit the data are reported in Table
\ref{table:spec}. The marginal discrepancy in flux is due to the difference in the absorption column density measured from the fits with the two spectral models. In the following we use $N_{\rm H}=0.6 \times 10^{22} {\rm cm}^{-2}$, as obtained from the fit with the {\sc compps} model. 
%Note, that any of the models used to fit the data required the addition of a soft thermal component ({\sc diskbb}). 
In Fig. \ref{fig:fig2} we show the unfolded spectrum and the best fit  {\sc compps}  model. The residuals from the fit are also shown. In this fit, the normalization of the ISGRI data was fixed to 1 as a reference, while the normalizations of the XRT and PCA data were $1.02\pm0.08$ and  $1.05\pm0.05$, respectively. A reasonable value for the normalization of the \rxte\ PCA could be obtained only once the Galactic ridge emission was taken into account in the background correction of this data.

\begin{figure}
%\vspace{0.25cm} 
\centering 
\centerline{\psfig{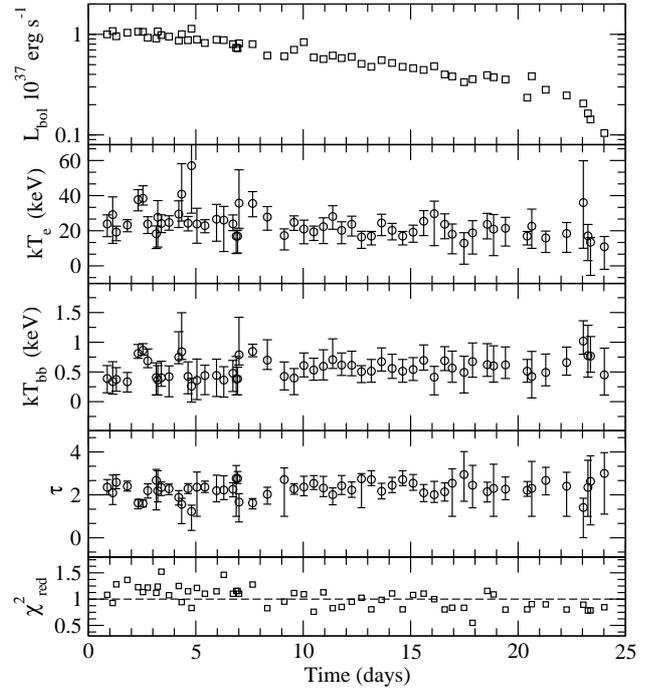}} 
\caption{The outburst evolution of the best-fit spectral parameters of
the {\compps} model using the \rxte/PCA/HEXTE data in the
2.5--200 keV range. Each spectrum corresponds to one \rxte\ observation
of \igr.  The bolometric luminosity is given in the energy range 0.5--200 keV assuming a source distance of 7 kpc.
}
\label{fig:fig3} 
\end{figure}

\begin{table}%[htb] 
\caption{\label{table:spec}Best-fit spectral parameters 
of the cutoff power-law and \compps\ models to the XRT/PCA/ISGRI data}
\centering
\begin{tabular}{lll} 
\hline 
& {\sc cutoffpl} & \compps \\
\hline 
\noalign{\smallskip}  
$N_{\rm H} (10^{22} {\rm cm}^{-2})$ & $0.96_{-0.09}^{+0.10}$ & $0.6_{-0.16}^{+0.14}$\\ 
$kT_{\rm e}$ (keV)& -- & $24.8^{+2.4}_{-2.3}$\\ 
$kT_{\rm bb}$ (keV)& -- & $0.59^{+0.08}_{-0.10}$\\ 
$\tau_{\rm T}$ & -- & $2.2^{+0.14}_{-0.17}$\\ 
$A_{\rm bb}$\tablefootmark{a} (km${^2}$)& -- & $260^{+80}_{-90}$ \\ 
$\cos \theta $ & -- & $0.53^{+0.09}_{-0.10}$\\
$\Gamma$ & $1.60^{+0.04}_{-0.04}$ & -- \\ 
$E_{\rm c}$ (keV) & $58.3^{+8.1}_{-6.5}$ & -- \\ 
$\chi^{2}/{\rm dof}$ & 112/140 & 106/138 \\
$F_{\rm bol}$\tablefootmark{b} ($10^{-9}$ erg cm$^{-2}$ s$^{-1}$) & 1.41$\pm$0.10 & 1.28$\pm$0.11\\
\noalign{\smallskip}  
\hline  
\noalign{\smallskip}  
\end{tabular}  
\tablefoot{ \tablefoottext{a}{Assuming a source distance of 7 kpc.}
\tablefoottext{b}{Unabsorbed flux in the 0.8--300 keV energy range.}
}
\label{tab:table1} 
\end{table}

\subsection{Spectral evolution during outburst} 
\label{sec:evolution} 
 
We analyzed all 58 \rxte/PCA/HEXTE (3--200 keV) spectra 
observed during the outburst. We excluded the last twelve \rxte\ pointings that were contaminated 
with the emission from a short weak outburst of XTE~J1751--305 (see Fig. \ref{fig:figima}, \ref{fig:fig1}, 
and Sect. \ref{sec:lc}).  In Fig. \ref{fig:fig3} we show the best-fit results using the thermal
Comptonization {\compps} model. The burst intervals have been removed. 
For the \compps\ model the  absorption column density and the inclination angle, $\theta$, were 
 fixed at the best-fit value found for the broad-band fit (see Table \ref{tab:table1}). The luminosity $L_{\rm bol}$
was calculated for a distance of 7 kpc (see Sect. \ref{sec:burst}) 
from the flux measured for the best-fit model in the energy range 0.5--200 keV.  
These results show that the decay of the outburst is marked by
a nearly constant plasma temperature, $kT_{\rm e}$, soft seed photons emission, $kT_{\rm bb}$, and optical depth, $\tau$. No statistically significant variations are measured.

A similar behaviour, althrough less evident due to less observational coverage during the outburst, 
was also observed in the case of the AMXP \igrone\ \citep{mfa05}. Note that  \igr\ and \igrone\  show
similar color variations during the outburst decay \citep{altamirano10,linares07}. 
The hardness-intensity diagram of \igr\ is shown in Fig. 2 of \citet{altamirano10}, 
and shows a stable (hard) color up to $\sim18$ days after the onset of the outburst, then the color softens within
the hard state. Our Fig. \ref{fig:fig3} also shows that the product $\tau \times kT_{\rm e}$, as measured during the outburst of \igr\ is stable during the outburst,  in agreement with observations
of other AMXPs \citep[e.g.,][]{gilfanov98,gp05,mfa05,p06,ip09}. 
Such behaviour is expected if the energy dissipation takes place in an accretion shock.  
The shock geometry can be approximated by a slab and the 
cooling of the hot electron gas is  determined
by the reprocessing of the hard X-ray radiation at the neutron star
surface (see \citealt{hm93,stern95,ps96,mbp01}). 
The temperature depends on the optical depth, but 
$\taut \times kT_{\rm e}$ is approximately constant. In
\igr\ we observe $kT_{\rm e}\times \tau_{\rm T}\approx 50$ keV (see
Fig. \ref{fig:fig3}) which is consistent with the values determined
for other AMXPs \citep{pg03,gp05,mfa05,mfb05,p06} as well as with the
theoretical models.

The spectra of AMXPs are very similar to those of the atoll sources at low luminosities
\citep{barret00}, where the X-rays are probably produced in the
boundary/spreading layer near the NS equator \citep{KW91,is99,sp06}.
Spectral similarities can be explained if in both types of sources, the energy
dissipation happens in the optically thin medium (i.e. accretion shock
and boundary/spreading layer) and the spectral properties are
determined solely by energy balance and feedback from the NS surface
which provides cooling in the form of soft photons.

\section{Timing characteristics}
\label{sec:tmchar}

We also studied the pulsed emission of \igr\ in the 3--300 keV
energy range using \rxte/PCA, HEXTE, and \Integ/ISGRI
data. As the flux decays over the course of the \Integ\
observation we selected only those \rxte\ observations that
overlap with the \Integ\ observation lasting from MJD
55090.926 to 55093.520 (UTC) (see Fig. \ref{fig:fig1}). In addition to the application of standard selectioncriteria in the screening process the 
\rxte/PCA data were further screened for bursts and detector
break-downs. The screening process yielded the following exposure
times for PCU detectors 0--4, respectively: 16.096 ks, 12.960 ks,
47.344 ks, 10.848 ks, and 13.216 ks.
For a discussion of timing behaviour over
the whole outburst we refer the reader to \citet{papitto10a,riggio10,ikp10}.

For HEXTE we only selected the on-source data streams from both
detector clusters. The dead time corrected exposure times in this case
are 30.2 and 16.0 ks for HEXTE{\footnote{HEXTE Cluster 0
operates in staring mode since July 13, 2006}} Cluster 0 and 1, respectively, 
for the period overlapping with the \Integ\ observation. For
the full 94041 observation the effective exposure times were 232.4 and
123.2 ks for Cluster 0 and 1, respectively.

%\clearpage

\begin{figure}[t] 
\centering 
\centerline{\psfig{figure=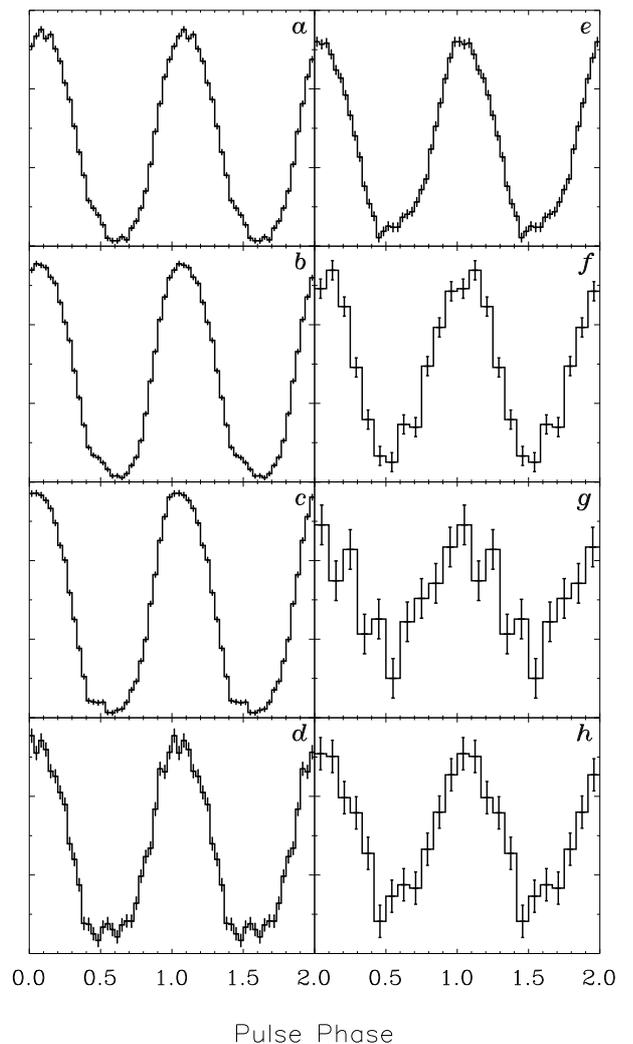,width=8.0cm,bbllx=150,bblly=150,bburx=440,bbury=650}} 
              {\caption[]{Pulse-profile collage of \igr\ using data
              from \rxte/PCA, HEXTE, and \Integ/ISGRI. Panels a--d (PCA) correspond to the
              energy intervals 1.7--4.2, 4.2--7.9, 7.9--16.2 and
              16.2--34.9 keV. Panels e--g show the HEXTE pulse
              profiles in the bands 15.6--31.0, 31.0--60.1 and
              60.1--123.9 keV (accumulated during the entire 94041
              observation period) in panel h the ISGRI
              profile 15--90 keV.  The error
              bars represent $1\sigma$ statistical errors. All
              profiles reach their maximum near phase $\sim 0.1$. 
              Note, the y-label are in units of counts per bin.
              \label{fig:profcol}}}
\end{figure}

We screened the ISGRI data for effects due to Earth radiation
belt passages or solar flare activity, none of the pointings
showed such activity. Time periods in which burst events occured
from any source in the ISGRI field of view were excluded from further
analysis. Furthermore, we selected only time stamps of events with
rise times between channels 7 and 90 \citep{lebr03}, detected in
non-noisy detector pixels which have an illumination factor of more
than 25\%.

The selected time stamps of all the instrument were then converted to
arrival times at the Solar System barycentre taking into account the
orbital motion of the spacecraft and correcting for acceleration
effects along the binary orbit. In this process we used the position
of the optical counterpart to \igr\ as reported by
\citet{torres09a,torres09b}.

\subsection{Pulse profiles and time lags}

Pulse phase folding of the barycentred arrival times using the
ephemeris given in \citet{papitto09} yielded the pulse-phase
distributions shown in Fig. \ref{fig:profcol}. In case of the PCA data
(collected in mode {\tt{E\_125us\_64M\_0\_1s}}) we added a time shift
of $+0.5 \times 122.07 \mu$s to the barycentred time stamps because
the times refered to the start of the time bin instead of the mid of
the bin. Panels a--d of Fig. \ref{fig:profcol} show the PCA pulse
profiles for the energy intervals, 1.7--4.2, 4.2--7.9, 7.9--16.2 and
16.2--34.9 keV, respectively. In panels e--g the HEXTE profiles
(accumulated during the full 94041 observation) are shown for the
bands 15.6--31.0, 31.0--60.1 and 60.1--123.9 keV, and panel h shows
the ISGRI profile for energies between 15 and 90 keV. Pulsed emission
has been detected up to $\sim 120$ keV using HEXTE data collected
during the full duration of observation 94041. 
The mutual alignment of the \rxte/PCA, HEXTE, and \Integ/ISGRI 
profiles within the equivalent energy bands is better than $50\mu$s.
The profiles are rather sinusoidal with the  amplitudes of harmonics not exceeding 
15\% of the fundamental \citep[see][]{papitto10a}. 
A small asymmetry could be caused by a strong deviation of emission pattern 
from the blackbody \citep{pg03,vp04} and/or the appearance at some phases 
of the secondary spot \citep{papitto10a} as seen in SAX J1808.4--3658 \citep{ip09,pia09}. 

\begin{figure} 
\centering 
\centerline{\epsfig{file=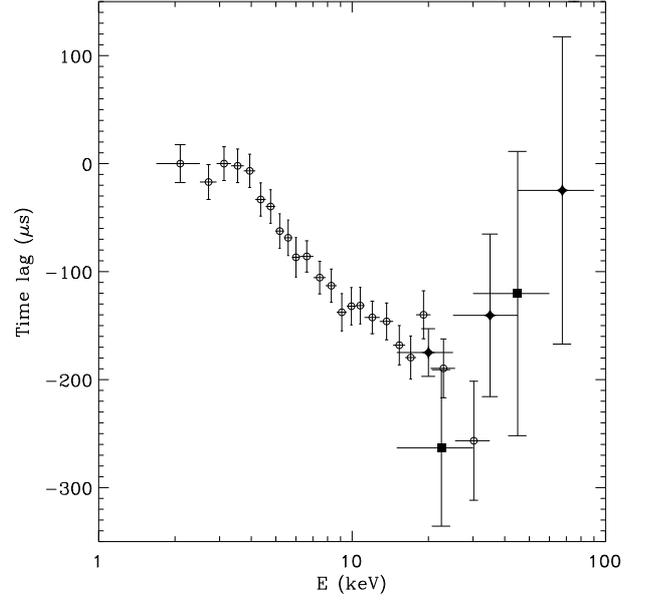,width=8.5cm} } 
\caption{Time lag  as a function of energy in the $\sim$2--100 keV
energy range combining \rxte/PCA (1.7--35 keV; open circles),
\rxte/HEXTE (15--90 keV; filled diamonds), and \Integ/ISGRI
(15--90 keV; filled squares) measurements.   
}
\label{fig:fig5} 
\end{figure}

We studied the global arrival times of the pulses as a function of
energy, and used the PCA profile for channel 1 ($\sim 1.69-2.51$ keV)
as a reference template. Cross-correlation of the pulse profiles
obtained for other energy bands with this template yielded the time
lags shown in Fig. \ref{fig:fig5}. Beyond $\sim 4$ keV a declining
trend sets in, meaning that the hard X-ray photons arrive earlier than the soft ones. 
%-- which does not seem to recover beyond $\sim 30$ keV given the poor statistics. 

The behaviour of the \igr\ time lags is similar to that observed in
XTE~J1751--305.  In these cases the low-energy pulses lag behind the
high-energy pulses (soft phase/time lags) monotonically increasing
with energy and saturating at about 10--20 keV \citep{gp05}.  The
saturation energy is, however, much smaller in \igrone\ \citep[see
e.g.][]{mfb05,ft07} and SAX~J1808.4--3658 \citep{CMT98,pg03,p06,ip09},
where the time lag as a function of energy breaks at about 7 keV.  For \igrone\  the soft lag even 
decreases beyond.
 
The lags are most probably due to different emission patterns of the blackbody 
and Comptonization components \citep{pg03,gp05,ip09} combined with the action of the 
Doppler effect. This is supported by the energy dependence of the lags, which 
grows until the contribution of the blackbody becomes negligible. 
At higher energies, the evolution of the soft lag is still further possible, because higher-energy 
photons suffer more scatterings resulting in variations of the 
emission pattern with energy \citep{pg03,vp04}. 
The break  in the lag spectrum  at  a higher energy 
might be a result of a higher seed blackbody temperature for Comptonization, 
which in turn implies a smaller spot area, and/or a smaller contribution 
of the blackbody to the total spectrum. Note that in XTE~J1751--305 \citep{gp05} and \igr\ the blackbody 
is hardly visible, while it is very apparent in the spectra of  
\igrone\ \citep{mfb05}.  and SAX~J1808.4--3658 \citep{gdb02,pg03,P08AIP}. 
A smaller blackbody area indicates either a larger magnetic field in the first two sources compared to \igrone,
or a different geometry (e.g. inclination), or a different optical depth of the Comptonizing plasma in the accretion shock. 
 
\begin{figure}[t] 
{\hspace{-0.25cm}\psfig{figure=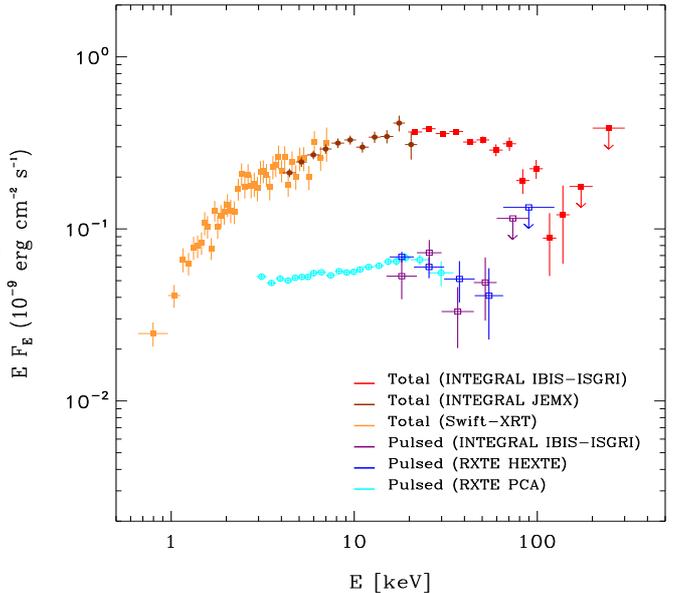,width=9.cm,height=8cm}}
{\caption[]{Unabsorbed total ($\sim$0.8--300 keV) and pulsed
($\sim$3--120 keV) unfolded spectra of \igr\ 
combining measurements from \rxte/PCA, HEXTE, and 
\Integ/ISGRI for the pulsed part and {\swift}/XRT, \Integ/JEM-X and ISGRI for the total part.  
\label{fig:fig6}} }
\end{figure} 

\subsection{Spectrum of the pulsed emission and pulsed fraction}

The pulsed spectrum and pulsed fraction (defined as pulsed flux/total
flux) as a function of energy provide important diagnostic parameters
for constraining the parameter space in theoretical modeling
\citep[see e.g.][]{vp04}. We derived the pulsed fluxes ($\sim$3--120
keV) from observations with the non-imaging \rxte/ PCA and HEXTE
instruments overlapping in time with the \Integ\ observation,
and the (imaging) ISGRI instrument. The total fluxes ($\sim$0.8--300
keV) have been determined from instruments with imaging
capabilities, namely {\swift}/XRT, and JEM-X and ISGRI
aboard \Integ\ (see also Sect. \ref{sec:spec}). 

Firstly, we derived the pulsed excess counts (=counts above DC level)
in a given energy band by fitting a truncated Fourier series, using
only the fundamental and two harmonics, to the measured pulse phase
distribution. For the PCA data ($\sim$3--35 keV band) the pulsed
excess counts were converted to photon flux values (ph
cm$^{-2}$s$^{-1}$\,keV$^{-1}$) in a forward spectral folding procedure
assuming an underlying power-law model taking into account the
different exposure times (see Sect. \ref{sec:tmchar}) and energy
responses of the five active PCU's.  We kept the absorbing hydrogen
column density fixed to $6.0\times 10^{21}$ cm$^{-2}$ (see Table
\ref{tab:table1}), although its precise value has very little impact on
the fit results because we consider only measurements with energies
above 3 keV. We obtained an unabsorbed 2--10 keV (pulsed) flux of
$8.09(6)\times 10^{-11}$ erg cm$^{-2}$s$^{-1}$ and a power-law index
of $1.824\pm0.004$ (reduced $\chi^2$ of the fit was 14.34/(21--2)),
indicating that the pulsed spectrum is 
softer than  the total spectrum (see Table \ref{tab:table1}).

For the HEXTE data ($\sim$15--120 keV) we employed an equivalent method:
the pulsed excess counts in a certain energy band (for those \rxte\
observations overlapping with the \Integ\ observation)
were divided by its effective sensitive area assuming a power-law
model with index $1.824$, taking into account the different energy
responses and dead-time corrected exposure times 
of the two detector clusters. Finally, the
pulsed ISGRI excess counts ($\sim$15--90 keV) of \igr\/ have been
converted to flux values adopting the method outlined in Section 3.4
of \citet{kuiper06}.

\begin{figure}[t] 
 {\hspace{-0.25cm}\psfig{figure=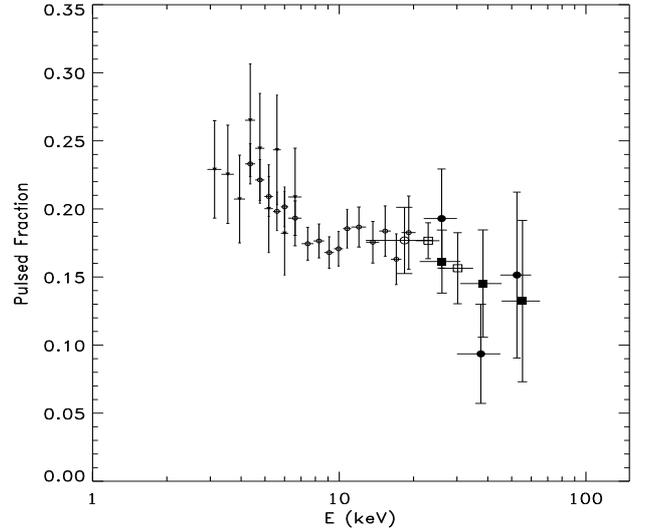,width=9.cm,height=7.5cm}}
 {\caption[]{The pulsed fraction (=pulsed flux/total flux) of \igr\
 based on the pulsed/total flux measurements from \swift/XRT,
 \rxte/PCA/HEXTE, and \Integ/JEM-X/ISGRI. 
 \label{fig:fig7}} }
\end{figure}

The (unabsorbed) PCA, HEXTE and ISGRI pulsed flux measurements are
shown in Fig. \ref{fig:fig6} along with the (unabsorbed) total flux
measurements.  If we compare the energy spectrum of the pulsed
emission of \igr\ and \igrone\ \citep[see][]{mfa05}, then we see a
significant difference: in the $E F_E$ spectral representation the
pulsed flux of \igr\ shows a cut-off near 20 keV, while that of
\igrone\ increases up to the end of the sensitivity window.  From the
pulsed and total flux measurements shown in Fig. \ref{fig:fig6} we can
directly derive the pulsed fraction as a function of energy which is
shown in Fig. \ref{fig:fig7}. 
The pulsed fraction decreases from $\sim$22\% at 3 keV to a constant pulsed fraction of $\sim$17--18\% 
between 7--30 keV, and then decreases again until $\sim$13\% at 60 keV. 
The decreasing part is very similar to what is observed in XTE J1751--305 \citep{gp05}, 
while in  \igrone\ the pulsed fraction has a minimum of about $\sim$6\% at 7 keV
 and grows again  to $\sim$12--20\% at 100 keV \citep{mfa05}. 

\section{Properties of the X-ray bursts} 
\label{sec:burst} 

\begin{table*} %[htb] 
\caption{Burst characteristics observed with \rxte/PCA and \Integ/JEM-X}
\centering   
\begin{tabular}{lccccccc} 
\hline 
 \# & Burst $T_{\rm start}$ & $F_{\rm pers, bol}$\tablefootmark{a}& 
 $F_{\rm peak}$\tablefootmark{b} & $f_{\rm b}$\tablefootmark{c} & $\tau_{\rm b}$\tablefootmark{d} & 
 $kT_{\rm bb,peak}$\tablefootmark{e} & $R_{\rm  bb,peak}$ \tablefootmark{f}\\ 
 & 2009 (UT) & $10^{-9}$ erg cm$^{-2}$ s$^{-1}$ & 
 $10^{-9}$erg cm$^{-2}$ s$^{-1}$ & $10^{-7}$erg cm$^{-2}$& s & keV& km\\
\hline 
\noalign{\smallskip}  
2&Sep 14 07:54:43 & $1.63\pm0.2$& $34.9\pm1.2$ &$3.20\pm0.03$ & $9.2\pm0.3$& $2.69\pm0.06$ & $5.5\pm0.3$\\ 
3&Sep 15 17:17:23 & $1.17\pm0.2$&  $33.0\pm1.2$ &$3.48\pm0.04$ & $10.5\pm0.4$& $2.59\pm0.05$ &$5.8\pm0.3$\\ 
4&Sep 17 06:33:27 &$1.48\pm0.2$ &  $38.5\pm1.2$ & $3.74\pm0.04$ & $9.0\pm0.3$ & $2.93\pm0.06$ & $4.8\pm0.2$\\ 
5&Sep 17 14:48:42 &$1.49\pm0.2$ &  $41.6\pm1.2$ & $3.52\pm0.04$ & $8.5\pm0.3$ & $2.99\pm0.06$ &$4.9\pm0.2$\\ 
6\tablefootmark{g}& Sep 18 08:09:29 &$1.28\pm0.2$& $45.0\pm9.0$& $3.30\pm0.6$& $6.6\pm1.9$& $3.60\pm0.9$ &$4.74\pm1.8$\\ 
7\tablefootmark{g}& Sep 19 02:28:20 &$1.31\pm0.2$ & $38.0\pm9.0$&$3.40\pm0.7$& $7.4\pm2.5$&  $3.40\pm0.9$&$4.74\pm1.8$\\ 
8\tablefootmark{g}& Sep 19 11:13:03 &$1.38\pm0.2$ & $36.6\pm10$&$3.90\pm0.7$& $7.9\pm2.9$&  $3.20\pm0.6$&$4.8\pm1.8$\\ 
9&Sep 20 14:50:31 &$1.36\pm0.2$ &  $42.4\pm1.2$ & $3.67\pm0.03$ & $8.6\pm0.3$ & $2.92\pm0.07$ &$5.1\pm0.3$\\ 
13&Sep 23 14:27:06 & $1.01\pm0.2$ & $48.5\pm1.2$ & $3.89\pm0.04$ &  $8.0\pm0.2$ & $3.26\pm0.1$ & $4.4\pm0.3$\\ 
14&Sep 24 18.39:13 & $0.95\pm0.2$ & $48.4\pm1.2$ & $3.84\pm0.04$ & $7.9\pm0.2$&  $3.15\pm0.09$&$5.4\pm0.3$\\ 
15&Sep 25 07:31:36 &$1.02\pm0.2$ &  $53.2\pm1.2$ & $3.94\pm0.02$ & $7.4\pm0.2$& $3.09\pm0.06$ &$5.2\pm0.2$\\ 
16&Sep 26 15:11:22 & $0.94\pm0.2$ &  $53.7\pm1.2$ & $4.18\pm0.05$ & $7.8\pm0.2$ & $2.98\pm0.07$&$5.6\pm0.3$\\ 
17&Sep 27 06:57:21&$0.88\pm0.2$ & $56.9\pm1.2$ & $4.21\pm0.04$ & $7.4\pm0.2$& $3.15\pm0.09$ & $5.1\pm0.3$\\
\noalign{\smallskip}  
\hline  
\noalign{\smallskip}  
\label{tab:burst} 
 \end{tabular} 
\tablefoot{ \tablefoottext{a}{Pre-burst unabsorbed flux in 0.8--200 keV energy range.}
\tablefoottext{b}{Burst peak flux in 0.1--40 keV energy band.}
\tablefoottext{c}{Burst fluence in 0.1--40 keV energy band.}
\tablefoottext{d}{Effective duration $\tau_{\rm b}=f_{\rm b}/F_{\rm peak}$.}
\tablefoottext{e}{Burst peak temperature.}
\tablefoottext{f}{Burst peak blackbody radius for distance of 7 kpc.}
 \tablefoottext{g}{Bursts detected with \Integ/JEM-X.}
}
\end{table*}
   
In Table \ref{tab:burst} we report the key measurable parameters for the
bursts observed from \igr.
Thermonuclear (type-I)
X-ray bursts are produced by unstable burning of accreted
% matter on the NS surface. The spectral shape in the energy
% range above a few keV can usually well be described 
matter on the NS surface. The spectrum from
a few keV to higher energies can usually be well described 
as a black-body with temperature $kT_{\rm bb}\approx$1--3 keV. 
The energy-dependent decay time of these bursts is
attributed to the cooling of the NS photosphere resulting in a gradual
softening of the burst spectrum \citep[see][for a review]{lewin93,sb06}. 

\begin{figure}[t] 
{\hspace{-0.25cm}\psfig{figure=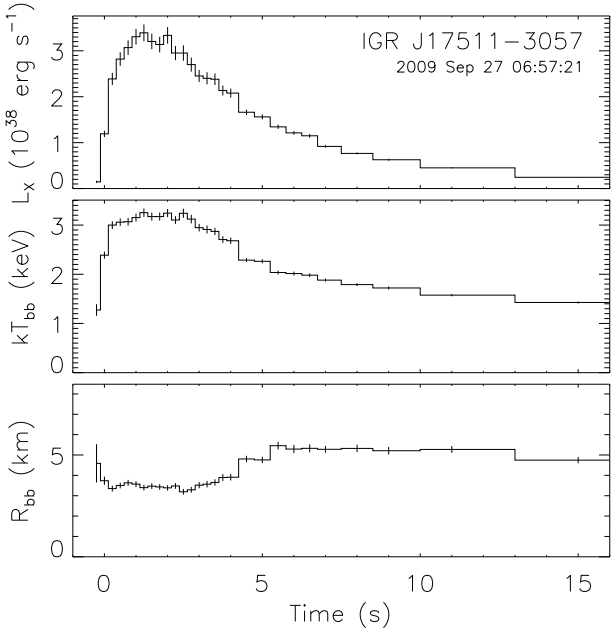,width=8.7cm}}
{\caption[]{Representative time-resolved spectroscopic results from the brightest burst seen by {\it RXTE}\/ from IGR J17511-3057. Shown is the inferred bolometric luminosity, from integrating the best-fit blackbody model ({\it top panel});  the blackbody temperature ({\it middle panel}); and the effective blackbody radius ({\it bottom panel}), assuming a distance of 7~kpc. 
\label{fig:specburst}}}
\end{figure}

We defined the burst start time as the time at which
the X-ray intensity of the source first exceeded 10\% of the burst peak
flux
(above the persistent intensity level). The time-resolved spectral
analysis of the 13 bursts (10 \rxte\, and 3 \Integ) was carried out by using
\rxte/PCA and \Integ/JEM-X data in the 2.5--20~keV and
3--20~keV bands, respectively. From these analyses we determined the
bursts' peak fluxes, temperatures, and radii (see Table
\ref{tab:burst}).  We fitted each burst spectra by a simple
photoelectrically-absorbed black-body,
\bb, model. The neutral absorption column density \nh\ was left free to
vary in the range (2.6-0.001)$10^{22}$ cm$^{-2}$ in all fits. However, we
checked that fixing \nh\ $=0.6\times10^{22}$ cm$^{-2}$ would not
significantly affect the results.
We extrapolated  the
unabsorbed fluxes to the 0.2--50~keV band by
generating dummy responses ({\sc xspec} version 12.6). This is
justifiable for the data since the black-body temperature is well
inside the spectral bandpass. 
The inferred \bb\ peak temperature, $kT_{\rm bb, peak}$, apparent \bb\
radius at 7~kpc (see below), $R_{\rm bb,peak}$,
and unabsorbed bolometric peak flux are also reported in Table
\ref{tab:burst}.  
In Fig. \ref{fig:specburst} we show representative time-resolved
spectroscopic parameters. 

We calculated the burst fluence, $f_{\rm b}$, by integrating the
flux over the burst duration. The effective
burst duration is $\tau_{\rm b}=f_{\rm b}/F_{\rm peak}$. 
All bursts are short, with $\tau_b$ in the range 7--10 s. The fluence and peak flux for the bursts
increased steadily with time, from $\sim 3.0\times10^{-7}$ to $4.2\times10^{-7}$ erg
cm$^{-2}$ s$^{-1}$ for the fluence, and from $\sim3.3\times10^{-9}$ to $5.7\times10^{-9}$
erg cm$^{-2}$ s$^{-1}$ for the peak flux. As a consequence, $\tau_b$
decreased steadily, and the rise time also decreased, from $\sim2$ to
$\sim1$ s. The brightest burst reached ($5.7\pm0.12$)$\times10^{-8}$
erg cm$^{-2}$ s$^{-1}$, and the minimum observed separation of the
bursts was 7.08 hr (see Table \ref{table:rec}). These results are in
agreement with those reported in \citet{altamirano10}.  Those authors also
reported the properties of burst oscillations detected at the pulse
frequency in all the bursts.
% and discussed in \citet{altamirano10}.

\begin{table}% [htb] 
\caption{Bursts recurrence times and effective exposure times.}
\centering 
\begin{tabular}{llcrrc} 
\hline 
 \# & Observatory & $ T_{\rm start}$ & $\Delta t_{\rm rec}$ (hr) & $T_{\rm exp}$ (hr) & $N$\\
\hline 
\noalign{\smallskip}  
1 & \swift & Sep 14 00:50:27& -- & -- & --\\ 
2 & \rxte & Sep 14 07:54:43 & 7.084 & 1.515 & --\\ 
3 & \rxte/\swift &Sep 15 17:17:23 & 33.395 & 10.454 & 3\\ 
4 & \rxte & Sep 17 06:33:27 & 37.253 & 10.88 & 3 \\ 
5 & \rxte & Sep 17 14:48:42 & 8.254 & 6.549& -- \\ 
6 & \Integ & Sep 18 08:09:29 & 17.331 & 7.859 & 1 \\ 
7 & \Integ & Sep 19 02:28:20 & 18.314 & 7.303 & 1 \\ 
8 & \Integ & Sep 19 11:13:03 & 8.745 & 8.226 & -- \\ 
9 & \rxte & Sep 20 14:50:31 & 27.639 & 5.056 & 2 \\ 
10& \xmm & Sep 21 01:12:21 &10.349 & 2.40 & -- \\ 
11& \xmm & Sep 21 12:34:01 &11.361 & 11.361& -- \\ 
12& \chandra & Sep 22 12:54:59 & 24.349 & 12.555 & 1 \\ 
13& \rxte & Sep 23 14:27:06 & 25.553 & 8.207& 1 \\ 
14& \rxte  & Sep 24 18.39:13 & 28.214 & 10.976 & 1 \\ 
15& \rxte & Sep 25 07:31:36 & 12.862 & 6.133 & --\\ 
16& \rxte & Sep 26 15:11:22 & 31.659 & 13.126 & 1 \\ 
17& \rxte & Sep 27 06:57:21& 15.770 & 4.985 & --\\ 
18& \swift & Sep 30 18:31:57 & 83.576 & 24.240& 3 \\
\noalign{\smallskip}  
\hline  
\noalign{\smallskip}  
\label{table:rec}
\end{tabular}  
\tablefoot{ The measured recurrence time between observed bursts is $\Delta
t_{\rm rec}$. The effective exposure time, $T_{\rm exp}$, takes into account all
the instruments observations. In Fig. \ref{fig:burst} the triangles corresponds to
$\Delta t_{\rm rec}/(N+1)$, with $N$ being the number of the expected missed bursts.
}
\end{table}

When a burst undergoes a photospheric-radius expansion (PRE), the
source distance can be determined based on the assumption that the
bolometric peak luminosity is saturated at the Eddington limit,
$L_{\rm Edd}$ \citep[e.g.,][]{lewin93,kuulkers03}. None of the observed
bursts exhibited PRE; thus, in these cases the peak luminosity is expected
to
have been sub-Eddington, leading to an upper limit on the distance. Assuming a bolometric peak luminosity equal
to the Eddington value for a He X-ray burst \citep[$L_{\rm
Edd}\approx\,3.8\times 10^{38}$ ergs$^{-1}$, as empirically derived
by][]{kuulkers03}, we obtain, using the brightest burst (see Table
\ref{tab:burst}), the source distance upper-limit of $d\lesssim7.5$
kpc. For comparison, the theoretical value of this upper limit 
distance \citep[e.g.,][]{lewin93} found by assuming a He atmosphere
and canonical NS parameters (1.4 solar mass and radius of 10 km), is
$\sim6.3$ kpc. Alternatively, assuming the peak luminosity to be the
Eddington luminosity for solar composition ($X=0.7$) implies 
a limit of $\approx4.8$ kpc. In the following, we consider
$d\approx$7~kpc to be a fiducial distance. 
At this distance, all bursts occurred at persistent luminosities 
between $(5.2-9.6)\times10^{36}$ erg s$^{-1}$ (see Table \ref{tab:burst}), 
or $\approx(1.4-2.5)\% L_{\rm Edd}$ using $L_{\rm Edd}\approx\,3.8\times 10^{38}$ erg s$^{-1}$.  The
local accretion rate per unit area for the pre-burst emission, $L_{\rm pers}$, is then
given by $ \dot m = L_{\rm pers} (1+z) (4\pi R^2(GM/R))^{-1}$,
i.e. $\dot m \sim (2.13-5.3)\times10^3 $g cm$^{-2}$ s$^{-1}$.  We use here the gravitational redshift $1+ z = 1.31$ for canonical NS mass, $M=1.4 M_{\odot}$, and radius, $R=10$ km. 

The observed energies of the bursts allow us to estimate the ignition
depths. The measured fluences of the bursts are
$f_{\rm b}=(3.2-4.2)\times 10^{-7}\ {\rm erg\ cm^{-2}}$, corresponding to a
net burst energy release $E_{\rm burst}=4\pi
d^2f_{\rm b}=(1.9-2.5)\times 10^{39}\ (d/7\ {\rm kpc})^2\ {\rm erg}$. The
ignition depth is given by $ y_{\rm ign} = E_{\rm burst} (1+z)(4\pi
R^2Q_{\rm nuc})^{-1}$, where the nuclear energy generated 
(assuming a mean hydrogen mass fraction at ignition $\langle X\rangle$) is
% value of $Q_{\rm nuc}\approx 1.6$ MeV
% corresponds to the nuclear-energy release per nucleon for complete
% burning of helium to iron group elements. Including hydrogen with a
% mass-weighted mean mass fraction $\langle X\rangle$ provides a value
% of
$Q_{\rm nuc}\approx 1.6+4\langle X\rangle$ MeV/nucleon
\citep[][and references therein]{galloway04}, including losses due to neutrino emission
following \citet{fujimoto87}. 
% For $\langle X\rangle=0.7$, the solar
% composition value, $Q_{\rm nuc}=4.4$ MeV/nucleon, and $y_{\rm ign}=(4.5-5.9)\times 10^7\ {\rm g\ cm^{-2}}$. 
% At the pre-burst
% accretion rate (see Table \ref{tab:burst}), the recurrence time,
% $\Delta t=(y_{\rm ign}/\dot m)(1+z)$, is 8.2--27.3 hr for helium and
% 3.1--10.0 hr  for hydrogen bursts.
Most burst sources accrete a mix of hydrogen
and helium, although the H-fraction at ignition may also be reduced by
steady burning between the bursts, i.e. $X\leq X_0$. Steady H-burning proceeds via the
hot-CNO cycle, which will exhaust the available hydrogen in approximately
11~hr~$(Z/0.02)^{-1}(X_0/0.7)$, where $Z$ is the mass fraction of CNO
nuclei, and $X_0$ the mass fraction of hydrogen in the accreted fuel
\citep[e.g.,][]{galloway04}. As for all bursts observed from \igr, the
bursts early in the outburst, when the recurrence time was as short as
7~hr, exhibit short profiles, characteristic of low H-fraction, and with
correspondingly high $\alpha$-values. In order to exhaust the accreted H in just
7~hr via steady burning, the accreted H-fraction must be substantially
below solar; we infer $X_0\leq 0.44(Z/0.02)$. Naturally, it is also
possible that the H-fraction is closer to the solar value, but that the
CNO metallicity is elevated.
For the inferred pure helium composition at ignition (i.e. $\langle
X\rangle=0$) the column depth varies little from burst to burst, in the
range
$y_{\rm ign}=(1.2-1.6)\times 10^{8}\ {\rm g\ cm^{-2}}$.

\begin{figure}[t] 
\centering
{\hspace{-0.25cm}\psfig{figure=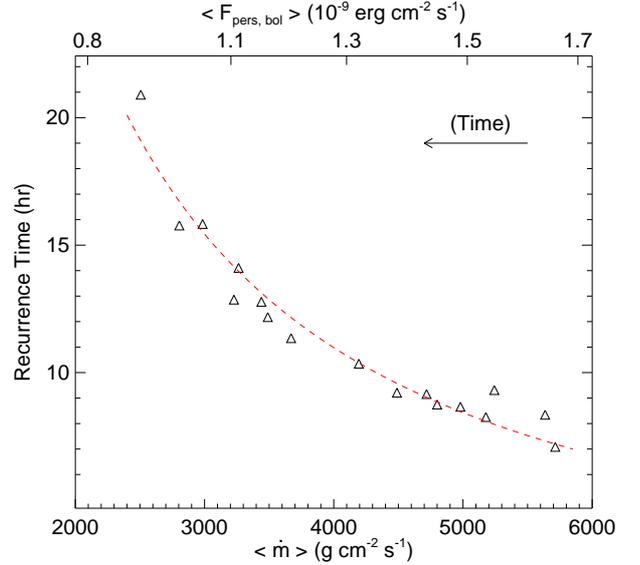,width=8.0cm}}
{\caption[]{The triangles in the figure represents the observed burst recurrence times (see Table \ref{table:rec}) shown as a function of local mass accretion rate (the corresponding flux is reported in the upper x-axis).  We also show the best fit power-law model. The recurrence time increased with time roughly as $\langle F_{\rm pers,bol}\rangle^{-1.1}$.

%The predicted burst recurrence time 
%as a function of local accretion rate (flux) (dots with error bars) assuming pure helium bursts for $\langle X\rangle = 0$ and the observed burst recurrence times (triangles). The latter  are divided for some missing %bursts by an integer number, $\Delta t_{\rm rec}/(N+1)$ (see Table \ref{table:rec}). 
%The triangle without the measured recurrence time is the \chandra\ reported burst. 
\label{fig:burst}} }
\end{figure}

Once steady H-burning exhausts
the accreted hydrogen at the base of the fuel layer, this process
no longer dominates heating in the layer, and ignition will
occur in a pure He-layer \citep[case 2 of][]{fujimoto81}. As the accretion
rate decreases through the decay of the outburst, we expect the burst recurrence
times to become longer, as is observed. We found that the recurrence time
increases roughly as $\langle F_{\rm pers,bol}\rangle^{-1.1}$ (Fig.
\ref{fig:burst}),
% Here $\langle\dot m \rangle $ is the averaged
% local mass accretion rate per unit area between the bursts.
where $\langle F_{\rm pers,bol} \rangle $ is the averaged
persistent flux between the bursts.
However, the expected increase
in the burst recurrence time in the pure He-ignition regime as a function
of decreasing $\dot{m}$
% (i.e. $\langle F_{\rm pers,bol}\rangle$)
is much steeper. For example, the curve for $X_0=0.3$
in Fig. 1 of \citet{galloway06} falls off roughly as $\dot{m}^{-3.5}$,
substantially steeper than measured in  \igr. What is also puzzling is
that the measured $\alpha$-value (the ratio of persistent to burst
luminosity) decreases with decreasing $\langle F_{\rm pers,bol}\rangle$.
As the burst recurrence time drops, the fuel layer should become
increasingly He-rich, so that a slight increase in $\alpha$ would be
expected.

% We also
% compare the expected recurrence time calculated as $\Delta t=(y_{\rm
% ign}/\langle\dot m \rangle)(1+z)$
% with the measured time interval, $\Delta
% t_{\rm rec}$, between bursts shown in Fig. \ref{fig:fig1}. 

This analysis relies on unambiguous measurement of the burst recurrence
time, for which we used all the bursts reported in the literature, 
with \swift\ \citep{bozzo09}, \xmm\ \citep{papitto10a},
and \chandra\ \citep{nowak09} in addition the \Integ\ and \rxte. 
% In Fig. \ref{fig:burst} we compare $\Delta t$ with its errors with $\Delta
% t_{\rm rec}$ (red triangles) assuming He bursts ($\langle X\rangle = 0$). 
% These derived recurrence times are independent of the assumed distance.
In some cases, the effective exposure time on source from 
burst to burst was shorter then the expected recurrence time
between bursts, so that we most likely miss one or more bursts between the observed one (see Table \ref{table:rec}). 
Therefore, we divided some measured times between bursts by an integer number, 
$\Delta t_{\rm rec}/(N+1)$ (see Table \ref{table:rec}), where 
$N$ is the number of missed bursts. 
We  verified that at the expected burst times we always had an observational data gap. 
% The measured recurrence times for $\langle X\rangle = 0$ are 
% significantly below the error bar at lower mass accretion rate. 
With the inclusion of observations for which there were no intervening data gaps
(e.g. with \xmm), we are confident that we have precisely inferred the
recurrence time between each pair of bursts listed in Table \ref{table:rec}.

The discrepancy between the expected and observed variation in recurrence
time is substantial, and in contrast to the (otherwise rather similar)
bursts observed at comparable accretion rates in SAX~J1808.4$-$3658
\cite[]{galloway06}.
As is often argued, it is possible that the variation in $\dot{m}$ is not
so large as is suggested by the decrease in $F_{\rm
pers,bol}$. If, as the X-ray intensity drops, we are seeing a
decreasing fraction of the accretion energy in the X-ray band, the actual
$\dot{m}$ could remain higher, which would be consistent
with both the moderate recurrence times and the decreasing $\alpha$
values. However, this would require a large change in the efficiency of
the $\dot{m}$ to $F_{\rm pers,bol}$
conversion. One way this could occur is that if the spectral energy
distribution changes such that the accretion flux is increasingly emitted
outside the band we are sensitive to. However, the combination of
instruments with both low- and high-energy sensitivity ({\it Swift} \&
{\it INTEGRAL}) makes this unlikely. \citet{thompson08}  found from a
detailed study of GS~1826$-$24 that the characteristic uncertainty that
might arise from this mechanism is 40\%. This is not sufficient to explain
the discrepancy in \igr.

\section{Summary} 
\label{sec:summary} 
   
We analyzed the simultaneous \Integ, \rxte, and \swift\ observations in order to study the
broad-band spectrum and timing behaviour of \igr. Using all
\rxte\ data we also studied the outburst profile. 
 The  broad-band average spectrum is well
  described by thermal Comptonization with an electron temperature of
  $\sim25$ keV and Thomson optical depth $\taut\sim2$ in a slab
  geometry. The object shows remarkable spectral stability during the
outburst marked by  constant plasma and seed photon temperature at a constant scattering
  optical depth.   We fitted the outburst profile with the exponential model and 
using the disk instability model  we inferred the outer disk radius to be $(4.8-5.4) \times 10^{10}$ cm. 

We showed that the coherent pulsation can be tracked with HEXTE and ISGRI instruments 
up to $\sim100$ keV. 
The pulsed fraction is shown to decrease from $\sim22\%$ at 3
  keV to a constant of $\sim$17--18\% between 7--30
  keV, and then to possibly decreasing further down to $\sim$13\% at 60 keV. This is similar to that observed in 
  XTE J1751--305, but markedly different from \igrone. 
  The nearly sinusoidal pulses show soft lags monotonically increasing with energy to about 0.2 ms at 10--20 keV,
  with some indications of a further decrease.  

Using all observations by \Integ, \rxte, \swift, \chandra, and \xmm\ we have collected 
the largest set of X-ray bursts observed from \igr, which allowed us to determine 
the recurrence time (accounting for the missed bursts) as a function of the accretion rate and the ignition depth. The short burst profiles indicate hydrogen-poor material at ignition, which suggests either that the accreted material is hydrogen-deficient, or that the CNO metallicity is up to a factor of 2 times solar. However, the variation of burst recurrence time as a function of $\dot{m}$ (inferred from the X-ray flux) is much smaller than predicted by helium-ignition models.

%\begin{figure}[t] 
%{\hspace{-0.25cm}\psfig{figure=fig8.ps,width=6.5cm,angle=-90}} 
%             {\caption[]{Companion mass M$_{\rm c}$ vs. radius R$_{\rm c}$ plane, showing 
%the Roche lobe constraints for burster X-ray millisecond pulsar (dashed line), for 
%$M_{\rm ns}=1.4 {\rm M}_{\odot}$. The circle indicate the binary system  inclination angle to be at the mean value of 60 deg.
%The figure also shows    the low mass regime degenerate brown dwarf 
%models incorporating different evolutions... 
%             \label{fig:fig8}} 
%              } 
%\end{figure}  

\begin{acknowledgements} 
MF thanks C. Winkler and the {\em INTEGRAL} staff for the rapid schedule of the \Integ\ observation of \igr\ shortly after the onset of its outburst. JP acknowledges financial support from the Academy of Finland grant 127512. 
\end{acknowledgements}

\end{document}